\documentstyle[11pt,epsfig]{article}
\begin{document} 
\begin{center}
{\huge \bf
            Freezing of Triangulations}\\
\vspace{1.5cm}
{\Large J.-P. Kownacki} \\
\vspace{1cm}
{\large
{ Laboratoire de Physique Th\'eorique et Mod\'elisation, \\
CNRS-Universit\'e de Cergy-Pontoise, UMR 8089\\
5, mail Gay-Lussac, Neuville-sur-Oise, 95031 Cergy-Pontoise Cedex, France\\
{\it E-mail : kownacki@ptm.u-cergy.fr}}}

\end{center}
\vspace{82pt}
{\abstract{
Zero temperature dynamics of two dimensional triangulations of a torus
with curvature energy is described. Numerical simulations strongly suggest that
the model get frozen in metastable states, made of topological defects on flat surfaces,
that group into clusters of same topological charge. It is conjectured that freezing
is related to high temperature structure of baby universes.}}
\phantom{.}
\vspace{8cm}
\begin{flushleft}
Published in Eur. Phys. J B 38, 485-494 (2004)
\end{flushleft}
\newpage
\section{Introduction}
Two dimensional dynamical triangulations models have been largely studied in the
context of euclidean quantum gravity over the past decades (for reviews, see \cite{rev_qg}).
These models allow
lattice regularization where the cut-off is the length of the links.
In view of 2d quantum gravity formulation, interesting features are essentially thermodynamical
limits and critical static properties, as they are key
features for continuum limit of lattice formulation. Great analytical progresses
have been made but numerical simulations are particulary suited to these models and allow
further investigations in remaining open problems. Square curvature terms in the
action/hamiltonian is such a problem that still lacks an analytical solution.

However, as random surfaces models, dynamical triangulations can
be used in several other domains like real membranes, foams, 2d-liquid, defects
on surfaces, {\it etc.} \cite{jeru89} \cite{bow97}.
Dynamical triangulations models have the nice feature to be very simple
to formulate - purely topological models - and to display very interesting physical
behaviours. 

Recently, Sherrington {\it et al.} have studied the dynamics of such a model
with curvature energy. They discovered behaviours characteristic of strong glass formers \cite{sherr}.
They underlined the role of topological constraints in this model without quench disorder,
and explained glassy behaviour as dynamics of topological defects at low temperature.
Their aim was to study a simple model of supercooled liquid, but their results are
not limited to it. In particular, dynamical properties are essential in numerical
simulations, even if static properties are studied - as in 2D quantum gravity - as
Monte-Carlo techniques consist in performing random walks in configuration space. In such
a context, a glassy behaviour with exponentially long equilibration times can mask
true equilibrium properties.

In this paper, the work described above is extended to zero temperature dynamics
of a square curvature energy model. In fact, zero temperature limit, 
where no thermally activated process is allowed, is crucial to understand, as
it gives more informations on the energy landscape and, thereby, on the whole
low temperature dynamics.

\section{The model - Static properties}
\subsection{Definition}
The set of all regular two dimensional triangulations \cite{DT} with $N$
vertices, toroidal topology and no boundaries is considered.
Regular means triangulations without tadpoles - vertex
connected to itself - and self-energy diagrams - two vertices connected by two
links. Moreover, each vertex is connected to at least three neighbours. This set is noted
$T^{torus}_{\mbox{\tiny N}}$. Topological constraints imply that the number of
vertices $N$, the number of links $N_l$
and the number of triangles $N_t$
are not independent. The first constraint sets the Euler characteristic
$\chi$ to zero for a torus.
$\chi$  is a topological invariant depending on the number of classes of nontrivial loops
on the surface - two for a torus. For a triangulation, it reads \cite{drou90}:
\begin{eqnarray}
   \chi = N - N_{l}+N_{t} \nonumber
\end{eqnarray}
so that, for a torus,
\begin{eqnarray}
N - N_{l}+N_{t} =0
\label{euler1}
\end{eqnarray}
The second constraint is local: each triangle is made of three links and, if there are
 no boundaries, each link belongs
to two triangles. So,
\begin{eqnarray}
    2  N_l  =  3 N_{t}
\label{euler2}
\end{eqnarray}
\noindent Combining relations (\ref{euler1}) and (\ref{euler2}) gives
\begin{eqnarray}
   N_l  =  3  N \ \ \ ,  \ \ N_t  = 2  N \nonumber
\end{eqnarray}

In the model described here, each triangulation ${\cal T}  \in T^{torus}_{\mbox{\tiny N}}$
is endowed with one - and only one - geometry. This is achieved by setting
the length of all links to unity and imposing on metric to be flat inside each triangle of $\cal{T}$.
Such geometries are called  piecewise linear manifolds.
It should be noticed that triangulations are not embedded
in any larger space, so that geometries are internal.

\begin{figure}[ht]
\begin{center}
\epsfig{file=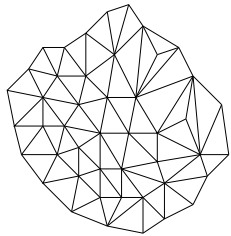,width=3cm,angle=0}
\end{center}
\caption{A triangulation.}
\label{triang}
\end{figure}

A random surfaces model is built from this ensemble of surfaces with fixed topology
but different geometries. The method consists to assign
to each triangulation/geometry an energy $E({\cal T})$ which governs dynamics and equilibrium.
At equilibrium,
each triangulation $\cal T$ is weighted by a Boltzman exponential
$\exp{\left( -\beta E({\cal T}) \right) }$
with  $\beta$
the inverse temperature. In the model, $E({\cal T})$ depends
on Gaussian - internal - curvature.
As geometry inside each triangle is flat,
curvature is concentrated at vertices. Following Regge calculus \cite{regge61},
an elementary surface
$ds_{i} =q_{i}$
is defined at each vertex
$(i)$ connected to $q_{i}$ neighbours (see figure \ref{triang}). Such a vertex is
called a $q_i$-vertex.

Local curvature at vertex $(i)$ is $q_{i} R_{i}$, with
\begin{eqnarray}
    R_i = \frac{(6-q_i)}{q_i} \nonumber
\end{eqnarray}
$R_i$  measures local deviation from flatness ($q_{i} = 6$ and $R_{i}=0$).
However,
$E({\cal T})$ cannot be the total curvature $\sum_{i} q_{i} R_{i}$ because
it is a topological
invariant, as stated by the Gauss-Bonnet theorem, and would be a constant in the model.
The choice for the energy is
\begin{eqnarray}
   E({\cal T})=\sum_{i} \ (q_{i}-6)^2
   \label{action}
\end{eqnarray}
It is not strictly speaking equal to the total square curvature
\begin{eqnarray}
\sum_{i} q_{i} R_{i}^{2} = \sum_{i} \ \frac{(q_{i}-6)^2}{q_{i}}
\label{R2}
\end{eqnarray}
but (\ref{action}) and (\ref{R2}) are expected to give similar results,
at least for small curvature, because
\begin{eqnarray}
\sum_{i} \ \frac{(q_{i}-6)^2}{q_{i}} = \frac{1}{6} \sum_{i} \ (q_{i}-6)^2 -
                 \frac{1}{36} \sum_{i} \ (q_{i}-6)^3  + \ldots \nonumber
\end{eqnarray}
In both cases, surfaces are flattened by curvature energy at low temperature.
However, expression (\ref{action}) is more symetric with respect to $q_{i}=6$.
Moreover, for surfaces containing only 5-, 6- and 7-vertices, the energy can
be rewritten
\begin{eqnarray}
       E({\cal T}) = n_{5}+n_{7} \nonumber
\end{eqnarray}
where $n_k$ is the number of $k$-vertices. As will be seen later,
these surfaces play a crucial role at low temperature.

Finally, the partition function of the model reads
\begin{eqnarray}
 Z_{N}(\beta) = \sum_{{\cal T} \in T^{torus}_{N}} \
              \frac{1}{\mbox{\small C}( {\cal T})}
                \ e^{- \beta E({\cal T})} \nonumber
\end{eqnarray}
where $C(\cal T)$ is a symetry factor needed to avoid  overcounting
symetric triangulations. For large $N$, $C({\cal T})=1$ for almost
all triangulations.

\subsection{Equilibrium Properties}
\begin{figure}[ht]
\begin{center}
\epsfig{file=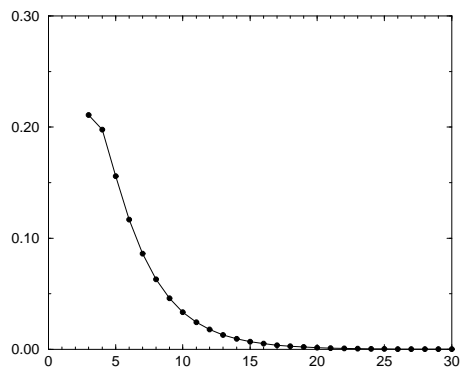,width=6cm,angle=0}
\end{center}
\caption{Equilibrium distribution $P(n)$ at $\beta=0$.}
\label{pn_theo}
\end{figure}

At infinite temperature ($\beta = 0$), the model has been solved \cite{tutte} 
\cite{brez78}. Typical
surfaces are characterized by hierarchical structures of baby
universes - bubbles growing on surfaces and linked
by very small necks. The distribution of their size is known \cite{jain92}.
As a consequence, typical surfaces are fractal with Haussdorf
dimension $d_{H}=4$ \cite{kaw93}.
The distribution of vertices with $n$ neighbours $P(n)$ has been calculated \cite{boula86}
(see figure \ref{pn_theo} ) and the variance is $\mu_2=10.5$, meaning
a wide range of vertices on the surfaces. Note that the model follows a modified 
Aboav's law \cite{godreche}.

At finite temperature, the model has not yet been solved and static properties have
been studied numerically by many
authors \cite{boula87} \cite{tsuda}. However, a very similar model
has been solved analytically, using matrix model techniques \cite{kaza96}.

\begin{figure}[ht]
\begin{center}
\epsfig{file=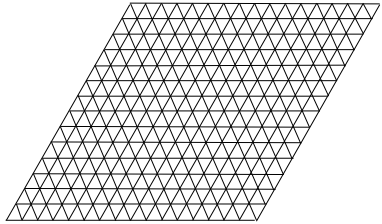,width=5cm,angle=0}
\end{center}
\caption{The hexagonal lattice.}
\label{hexa}
\end{figure}

When temperature is lowered, curvature energy favours flattening of surfaces.
The ground state of the model is the hexagonal lattice with $q_i=6$ at
each vertex (figure \ref{hexa}).
It should be emphasized that the ground state does not depend on the precise
form of the energy. In fact,  $E(\cal T)$ can be rewritten
\begin{eqnarray}
E(\cal T) &=& \ \sum_i \ \left( {q_i}^2 +36 -12 q_i \right) \nonumber \\
          &=& \ \sum_i \ \left( {q_i}^2 \right) + 36 N - 24 N_l \nonumber \\
          &=& \ \sum_i \ \left( {q_i}^2 \right) - 36 N \nonumber
\end{eqnarray}
so that, at given $N$, $E(\cal T)$ is equivalent to $\sum_i
 (q_i - q_o)^2$, whatever the value of $q_o$.
On the other hand, the mean neighbours number
\begin{eqnarray}
<q_i> &=& \frac{1}{N} \ \sum_i q_i \nonumber \\
&=& \frac{1}{N} \ 2 \ N_{l} \nonumber \\
&=& 6 \nonumber
\end{eqnarray}
is completely determined by
topological constraints (\ref{euler1}) and (\ref{euler2}).
So, the ground state is the configuration that minimizes $\sum_{i} q_{i}^{2}$ with
$<q_{i}> =6$.

\begin{figure}[ht]
\begin{center}
\epsfig{file=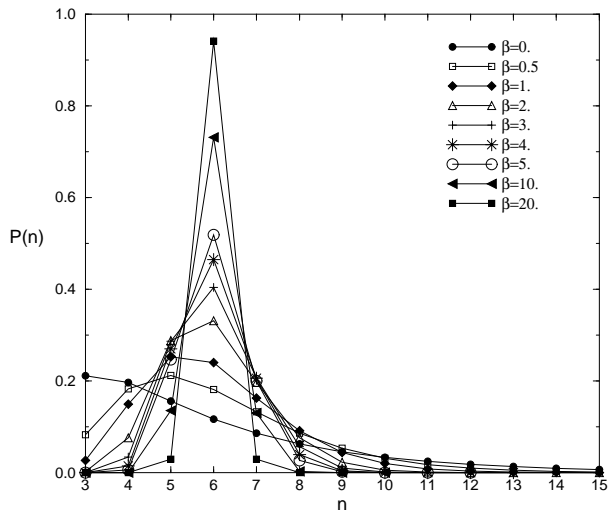,width=8cm,angle=0}
\end{center}
\caption{Equilibrium distribution $P(n)$ at various $\beta$, for $N=65025$.}
\label{Pn_beta}
\end{figure}

Here are summarized results obtained by numerical simulations, done as preliminary
work in this paper, for sizes up to $N=65025$. Starting from hexagonal lattices, the system
is initially equilibrated with typically 1000 sweeps, and simulations are carried with
about 3000 sweeps. For details
about the rules of simulations, see paragraph "dynamics" below.
As can be seen in figure \ref{Pn_beta}, the distribution $P(n)$ is more and more
peaked at $n=6$ as temperature is lowered,
meaning that surfaces contain essentially 6-vertices. Then, 5- and 7-vertices
can be seen as topological defects
in flat surfaces, whose proportion vanishes as temperature is lowered. It should be
noticed that, when only 5- , 6- and 7-vertices survive, $n_5 = n_7$. 
It is a consequence of topological
constraints (\ref{euler1}) and (\ref{euler2}), because
\begin{eqnarray}
N &=& n_{5}+n_{6}+n_{7} \nonumber \\
2 \ N_{l} &=& 5 \ n_{5} + 6 \ n_{6} + 7 \ n_{7} \nonumber
\end{eqnarray}
so that
\begin{eqnarray}
5 \ n_{5} + 6 \ n_{6} + 7 \ n_{7} = 6 \ (n_{5}+n_{6}+n_{7}) \nonumber
\end{eqnarray}
hence
\begin{eqnarray}
n_{5} -  n_{7} = 0 \nonumber
\end{eqnarray}
Baby universes also disappear at low temperature, as, otherwise, they would increase 
curvature energy.
However, as can be seen in simulations, there is no sign of phase transition between
high and low temperature phases. This is also suggested by dimensional analysis,
 as square curvature is not
relevant in the infrared limit. Moreover, the same conclusion arises in the
matrix model mentioned above.

\section{Dynamics}
Although static properties are rather simple, dynamics of the model reveals
a very rich structure.
\subsection{Finite temperature}
\subsubsection{The rules}
\begin{figure}[ht]
\begin{center}
\epsfig{file=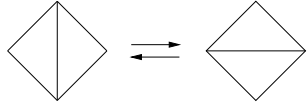,width=4cm,angle=0}
\end{center}
\caption{$T_1$ move.}
\label{T1mv}
\end{figure}

The rules of dynamical evolution are based on the energy (\ref{action}) and use
so-called Alexander moves \cite{alex}, more precisely $T_1$ moves, which consist in flipping
a link in a
triangulation (figure \ref{T1mv}). Topology and number of vertices $N$ are conserved
during evolution. It has been shown that
$T_1$ moves allow
ergodic explorations of the space of triangulations \cite{boula86}.

One elementary Monte-Carlo step of evolution is done as follows:\\
\noindent - Random choice of a link in triangulation $\cal T$;\\
\noindent - Flipping of the link with probability (Glauber type)
\begin{eqnarray}
 w({\cal T} \ \rightarrow {\cal T}^{'}) =  \ \frac{1}{1 + e^{\beta \Delta E}} \nonumber
\end{eqnarray}
where $\Delta E = E({\cal T}^{'}) - E({\cal T})$. The
proposed triangulation
${\cal T}^{'}$ is rejected if it  contains vertices with $q_i < 3$, self-energy diagrams or
tadpoles. At zero temperature ($\beta = + \infty$), the transition probability becomes
 \begin{eqnarray}
 w({\cal T} \ \rightarrow {\cal T}^{'}) &=& 0 \ \ \ \ \mbox{if $\Delta E > 0$}
 \nonumber \\
                             &=& \frac{1}{2} \ \ \ \  \mbox{if $\Delta E = 0$
} \nonumber \\
                             &=& 1 \ \ \ \ \mbox{if $\Delta E < 0$} \nonumber
\end{eqnarray}
In the following, times are expressed in unit of sweeps, $i.e$ $N_l$ successive
elementary Monte-Carlo steps.
\subsubsection{Quench to low temperature}
Sherrington {\it et al.} \cite{sherr} have studied the behaviour of the system after a
quench
from initial infinite temperature (disordered) state to low - but finite - temperature.
Their main  result is the emergence of glassy behaviour. It may looks rather
surprising as the model
presents neither frustration nor quenched disorder. There are
two regimes in the evolution, depending on the temperature of the quench. For high
temperatures, equilibrium is reached after a rapid relaxation, with equilibration
times independent of the temperature. \\
On the other hand, for low temperatures, the system evolves in two steps. First,
a rapid relaxation that corresponds to annihilation of all topological defects
except 5- and 7-vertices, with fast decay of energy. Then,
the system becomes glassy with
equilibration times exponentially growing with inverse temperature (Arrhenius behaviour).
In this second period, the energy reaches
a plateau, while
two times correlation functions of the curvature (local energy)
develop shoulders and loose invariance in time translation. Glassy behaviour is
the result of creations, annihilations and diffusions of 5- and 7-vertices. Simplified
reaction-diffusion models give similar results \cite{drou_comm}. The authors conclude that the model
displays properties characterizing strong glass forming systems \cite{kob99}.

\section{Zero temperature and freezing}
The particular case of deep quench to zero temperature with infinite cooling rate \cite{sas98}
is now described.
This is the original contribution of this paper.
\subsection{Energy}
\begin{figure*}[ht]
\begin{center}
\epsfig{file=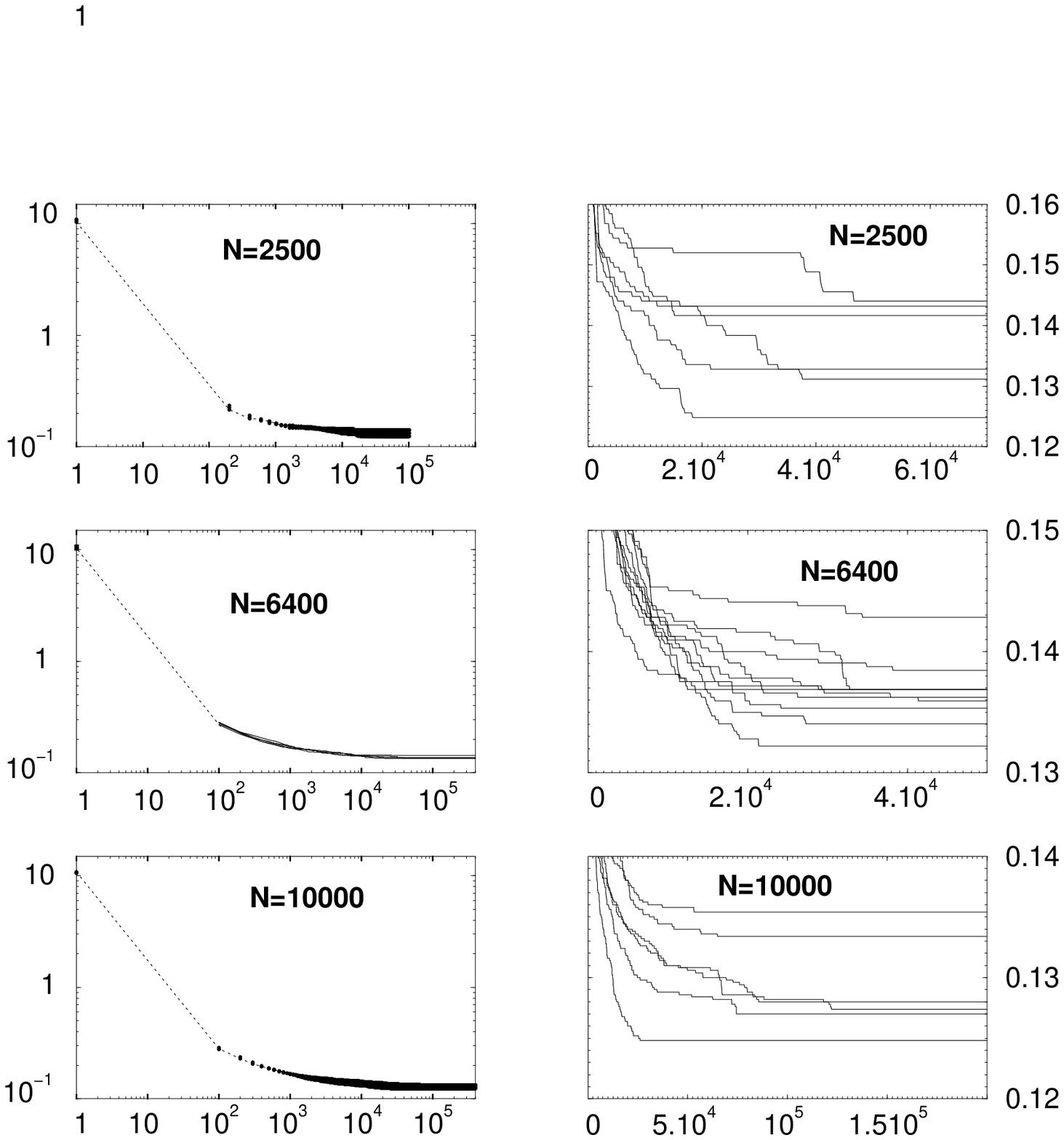,width=11cm,angle=0}
\end{center}
\caption{Evolution of energy after a quench to zero temperature, for $N=2500, 6400$ and 10000. 
Each curve corresponds to one run.}
\label{ener_evol}
\end{figure*}

The evolution of the energy $E(t)$ after a quench from infinite
temperature state to zero temperature is plotted in figure \ref{ener_evol} for several sizes of
the system.
Immediately after the quench, there is a very fast decay of the energy, followed by
a slowing down period where $E(t)$ reaches a serie of plateaux. At the beginning, this looks like
stairs. But as time goes, the plateaux become broader and broader and eventually ``infinite''. It
strongly
depends on the size of the system. However, simple contemplation of the graph of $E(t)$
is not enough to know whether energy really stops or not. $E(t)$ could as well decrease to
lower plateaux,
and eventually to zero, after very long times unreachable by reasonable computer simulations.
However,
as will be shown in the following, it is very likely that, after a deep quench, the system
gets trapped in metastable states. 

\begin{figure*}[ht]
\begin{center}
\epsfig{file=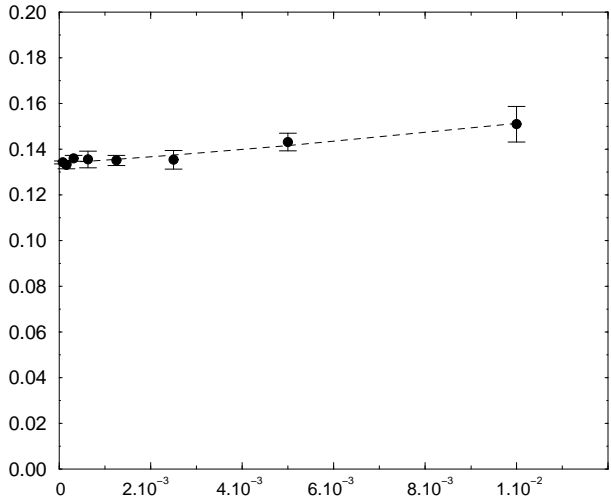,width=8cm,angle=0}
\end{center}
\caption{Frozen energy per site $E_{f}/N$, $versus$ $\frac{1}{N}$. The dashed line indicates
a power law fit.}
\label{ener_quench}
\end{figure*}

The final value reached by the energy follows a sharp distribution centered on a mean value
$E_f$, according to initial disordered states. For each data point in figure \ref{ener_quench},
16 independent runs are performed with 1000 heating
sweeps at infinite temperature followed by sufficient sweeps at zero temperature
(up to 5.$10^{5}$ sweeps for $N=12500$) - depending on the time
needed by energy to get frozen.
The first important point is that $E_f$ is strictly greater than zero: the  system is trapped
in metastable states and
never goes to its ground state. In other words, the energy freezes, in a time that
depends on the size of the system.
This can be seen by looking at the evolution of $P(n)$, which shows a very fast invasion 
of surfaces by 6-vertices. Both descriptions - in terms of energy and in terms of defects -
are equivalent as energy
and number of defects $n_{defects}$ are simply related, if defects are 5- and 7-vertices only: 
\begin{eqnarray}
E &=& n_{defects}  \nonumber
\end{eqnarray}

The second important point is that  $E_f/N$ is constant for large $N$,
as can be seen in figure \ref{ener_quench}. A power law fit gives a limiting value
$\lim_{N \to \infty} E_f/N = 0.1342(4)$. So, mean frozen energy per site
is a thermodynamical characteristic of the model.
Alternatively, a finite fraction - $\sim 13 \%$ -
of defects remains in frozen states.

\subsection{$T_1$ Moves}
\begin{figure*}[ht]
\begin{center}
\epsfig{file=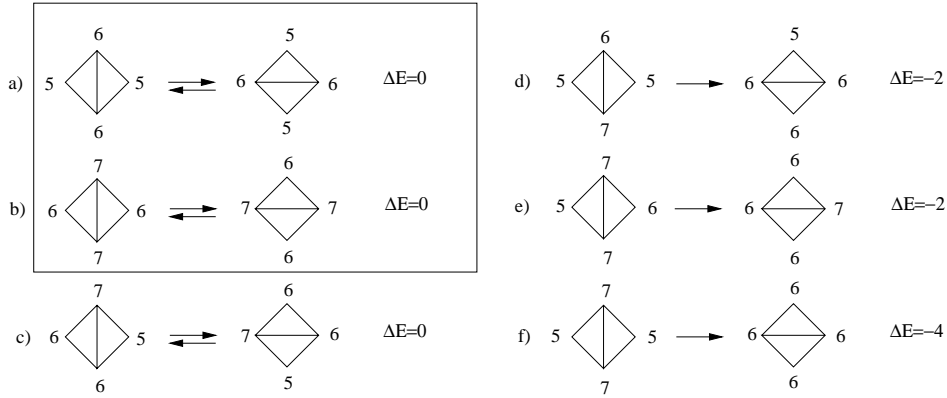,width=13cm,angle=0}
\end{center}
\caption{$T_1$ moves allowed at zero temperature. When there are no dipoles (5-7), only the two
first moves (in the box) can occur.}
\label{list_flipTOT}
\end{figure*}

\begin{figure*}[ht]
\begin{center}
\epsfig{file=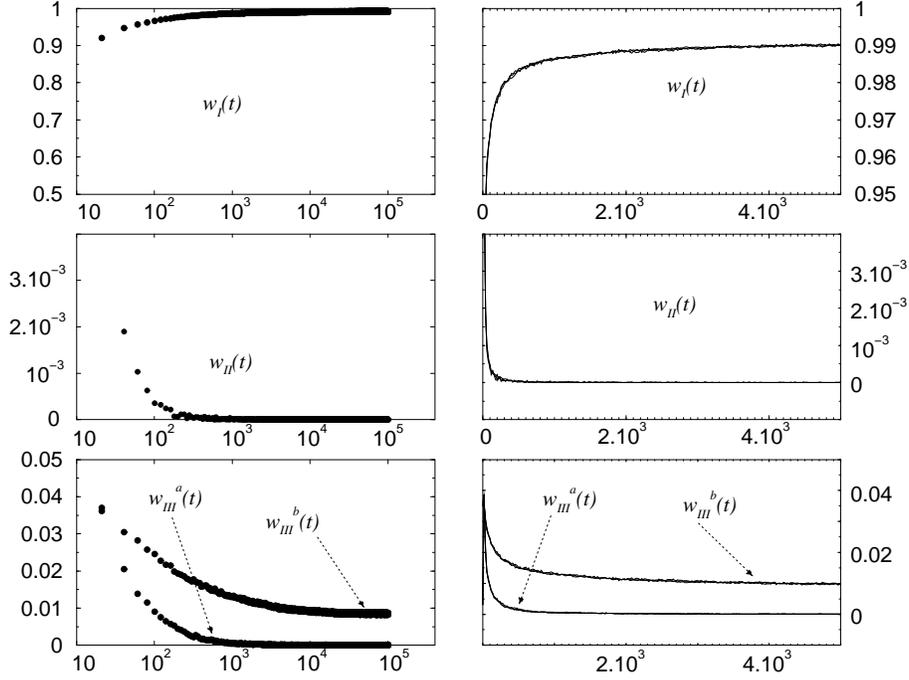,width=12cm,angle=0}
\end{center}
\caption{Probabilities of flips $w_{I}$, $w_{II}$,
 $w_{III}^{a}$ and $w_{III}^{b}$ {\it versus} time for $N$=2500, 4900 and 10000. Curves corresponding to 
different sizes are not distinguishable.}
\label{proba_flips}
\end{figure*}

\begin{figure}[ht]
\begin{center}
\epsfig{file=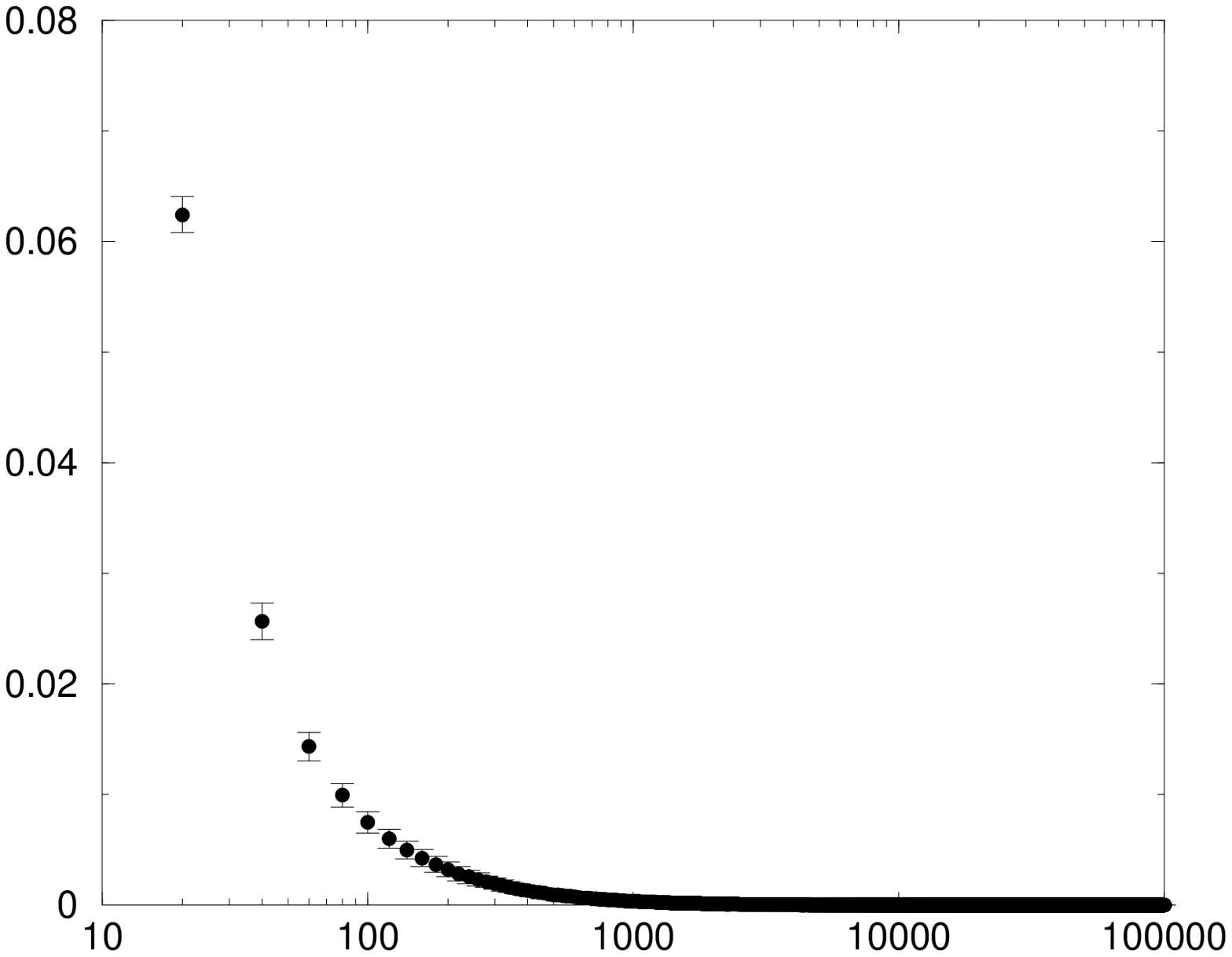,width=8cm,angle=0}
\end{center}
\caption{Fraction of dipoles {\it versus} time for $N=2500$, averaged on 576 independent runs.}
\label{dipoles}
\end{figure}

\begin{figure}[ht]
\begin{center}
\epsfig{file=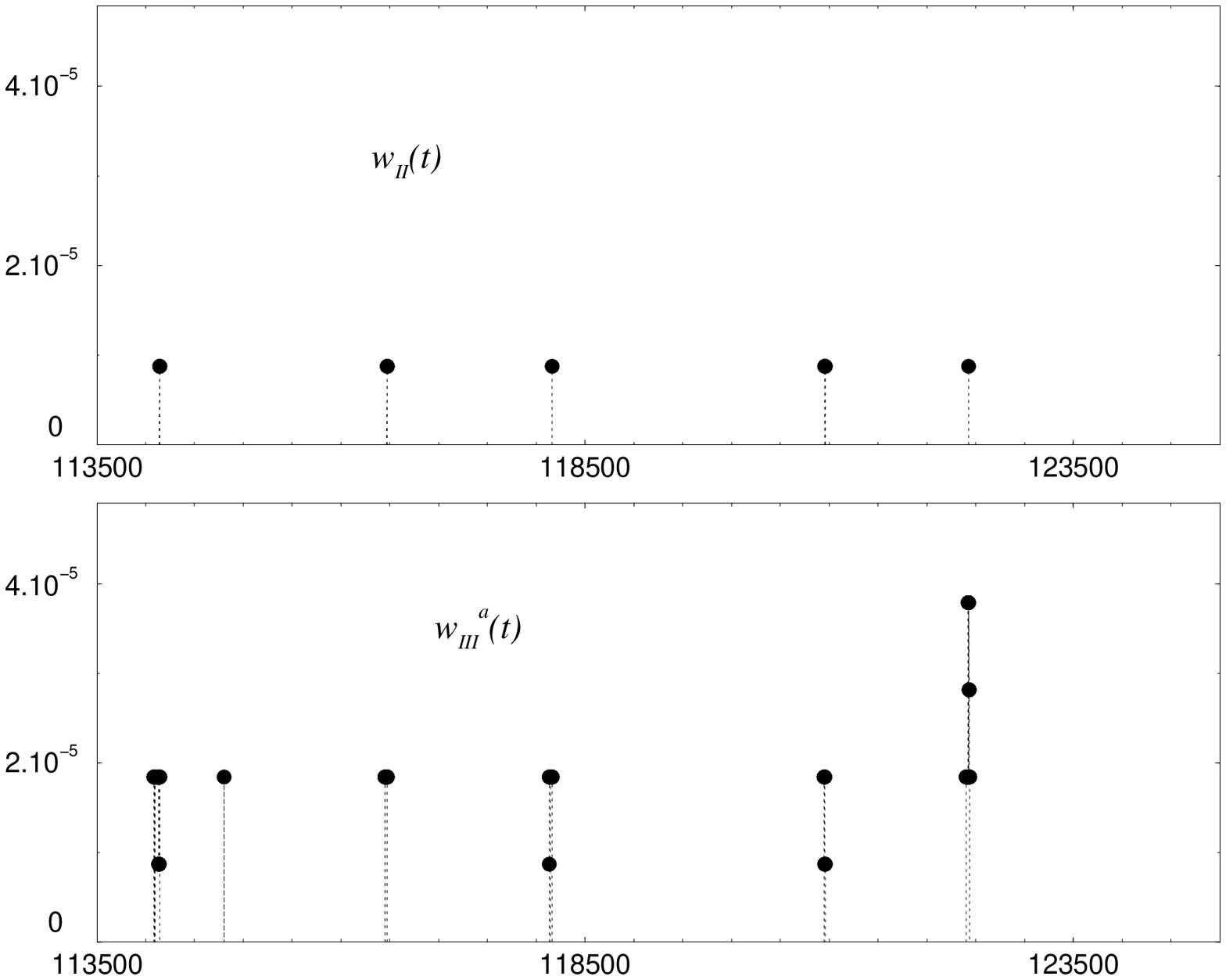,width=8cm,angle=0}
\end{center}
\caption{Detail of $w_{II}$ and $w_{III}^{a}$ for one run and N=4900.}
\label{detail}
\end{figure}

At zero temperature, besides 6-vertices, only 5- and 7-vertices survive.
Evolution of the system can be seen
as dynamics of such defects through
the six possible $T_1$ moves shown in figure \ref{list_flipTOT}.
Moves $d)$, $e)$ and $f)$ have
negative energy balance ($\Delta E < 0$) and correspond to annihilation of defects. 
The others - $a)$,
$b)$ and $c)$ -
can be viewed as diffusion of defects on the surface. In this second category, 
moves $a)$ and $b)$ - called flippers in the following - perform
local rotations of 5-5 or 7-7 pairs. On the other hand,
move $c)$ is a true diffusion of a dipole - $i.e$ a pair
of 5-7. In order to understand the evolution of the system after a deep quench, it is important
to know the possible ways of instantaneous evolution of the system at a given time.
This depends on the repartition of $T_1$ moves and gives a partial view on the energy landscape 
\cite{bro00}.
More precisely, at a given time $t$, it is possible to
group the links of the surface into three categories:
the first one $(I)$ contains links that cannot be moved because energy would be increased; the
 second one $(II)$
contains links whose moves make energy strictly lower if they are flipped; and the third one 
$(III)$ contains links that let
energy unchanged. Two subcategories of $(III)$ can be defined: $(IIIa)$ contains links 
in $(III)$ correponding to dipole diffusion,
and $(IIIb)$ contains links in $(III)$ corresponding to flippers.
Proportions of links in $(I)$, $(II)$, $(IIIa)$ and
$(IIIb)$ at time $t$ after
a deep quench  are respectively written $w_I(t)$, $w_{II}(t)$, $w_{IIIa}(t)$ and
$w_{IIIb}(t)$. They are plotted for different sizes in figure \ref{proba_flips}, obtained
by averaging about 8 independent runs for each size.
It is
not surprising to see that $w_I(t)$ grows rapidly and, conversely, that $w_{II}(t)$ decreases 
rapidly:
 this is simply the
evolution toward the ground state - where all links are in $(I)$ - as expected for a system
 at zero temperature.
However, after this first rapid evolution - corresponding to rapid decay of energy - there
is a slowing down period where $w_{II}(t)$ is very small,
meaning that
the system can hardly lower its energy. The reason is that the system is almost unable to found 
local configurations that could lead to $d)$, $e)$ or $f)$ moves. On the other hand,  $w_{IIIa}(t)$ 
is strictly greater
than zero, so that diffusions
of dipoles still occur.  Through these diffusions, dipoles can eventually get close to other defects
 and give local
configurations of type $(II)$. However, creation of such configurations by this mechanism is rather
unlikely because
dipoles are rare in this phase (see figure \ref{dipoles}).
As a consequence, the dynamics is very slow. This mechanism is
confirmed by carefully looking
at $w_{II}(t)$ and $w_{IIIa}(t)$ simultaneously (see figure \ref{detail}).
Small peaks of $w_{II}(t)$  occur immediately after
small peaks of $w_{IIIa}(t)$, meaning that diffusion of pairs is the only way to create local
configurations leading to energy decrease. So, the role of $(IIIa)$ is crucial in the slowing
down period.

However, after some time, $w_{II}$ and $w_{IIIa}$ become equal to zero: this
coincides with energy freezing and removal of dipoles. But the dynamics is not really
stopped, because during the slowing down period described above, $w_{IIIb}$ reaches a
plateau with small but finite value - independant of the size -  so that flippers still occur even after
removal of dipoles and energy freezing.

At this point, the problem is to know whether flippers eventually
allow diffusion of defects, through some collective moves, that could lead, after a long time, to creation
of dipoles. In this case, energy could decrease again, and freezing of energy would be an illusion due to short
time measurements.

\subsection{Geometric description}
\begin{figure*}[ht]
\begin{center}
\epsfig{file=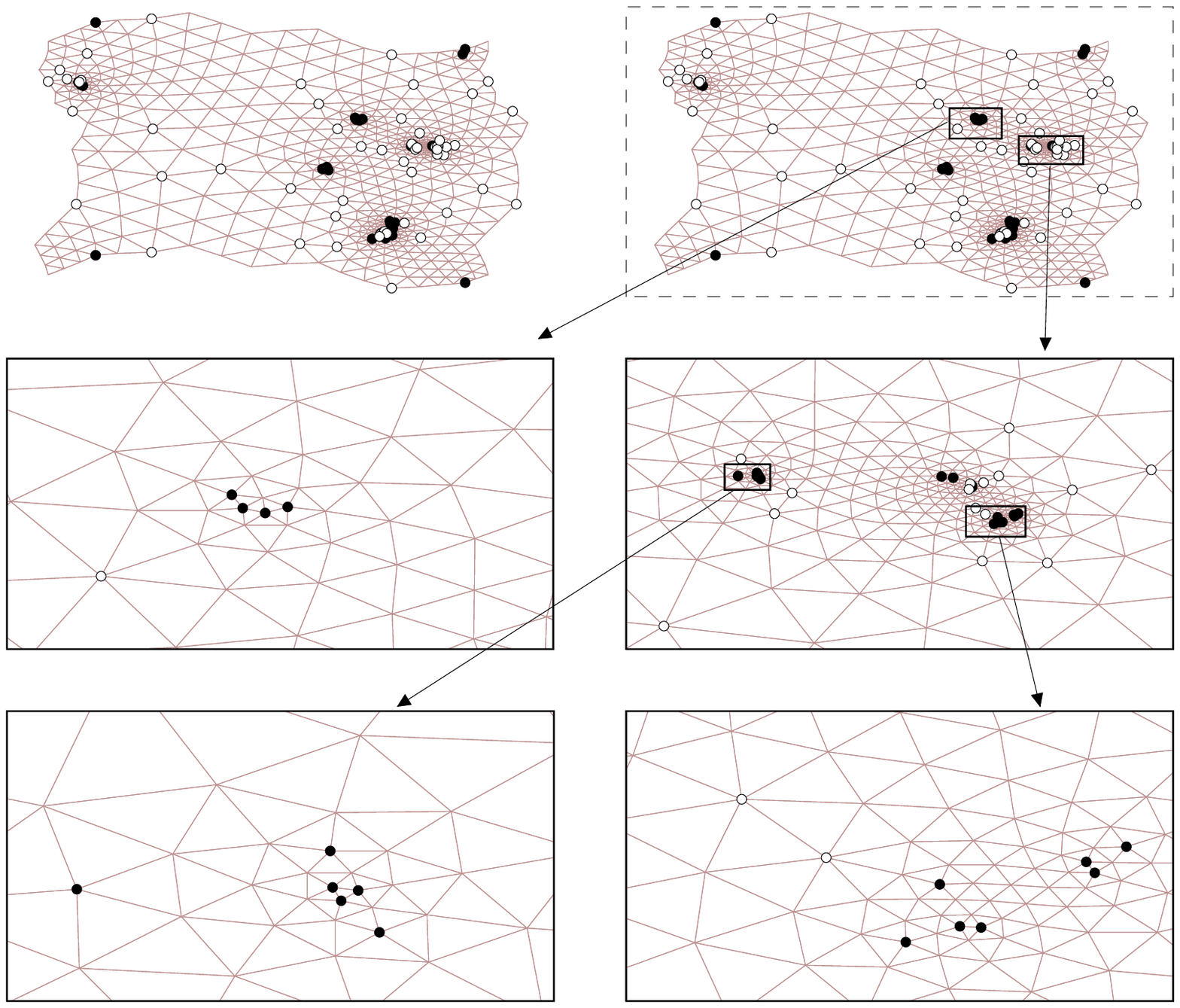,width=13cm,angle=0}
\end{center}
\caption{A frozen surface with $N=900$. 5-vertices are represented by $\bullet$ and
7-vertices by $\circ$. All others vertices are 6-vertices. Arrows indicate
zooms of regions in the boxes.}
\label{snapshots}
\end{figure*}

\begin{figure}[ht]
\begin{center}
\epsfig{file=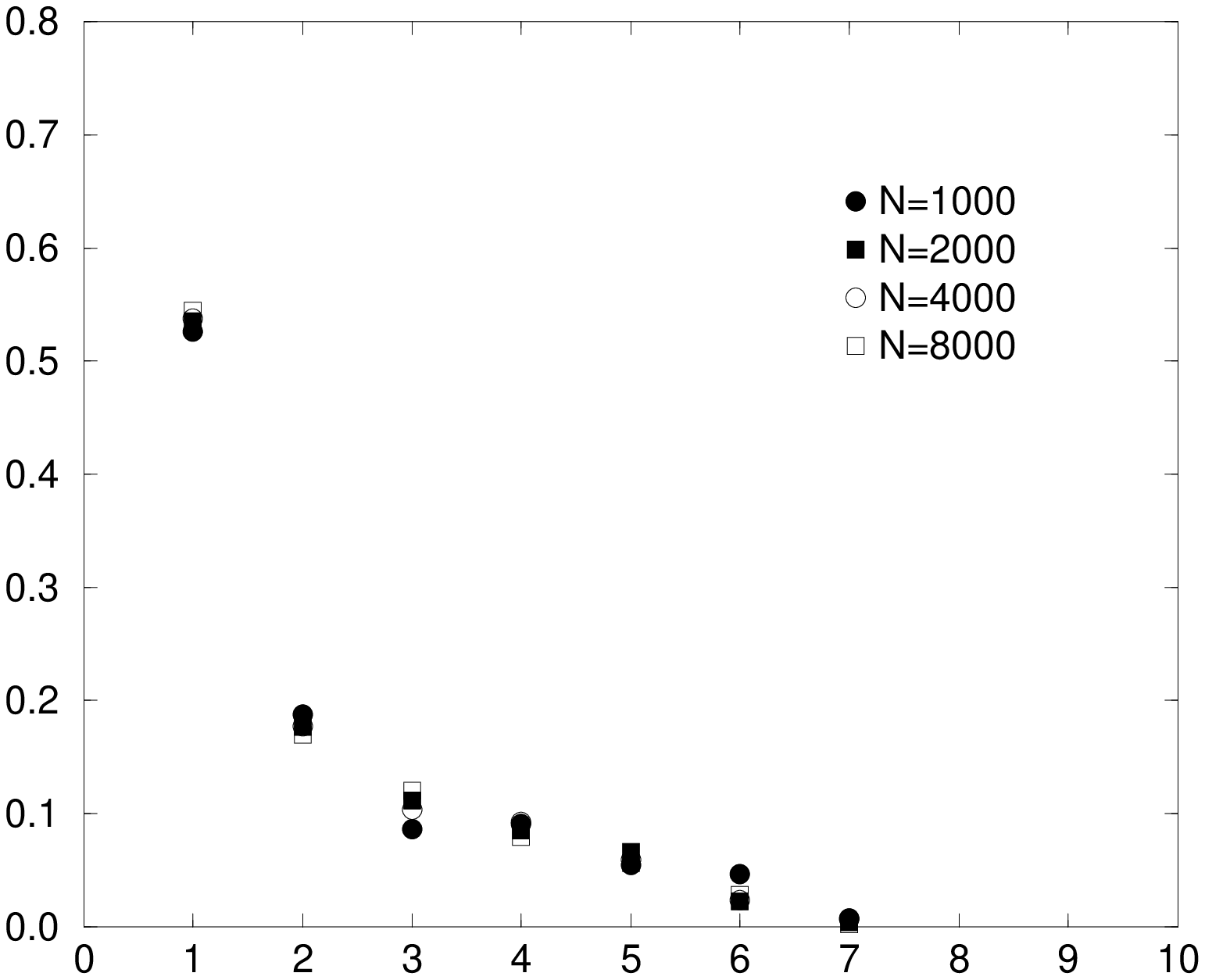,width=6.5cm,angle=0}
\end{center}
\caption{Size distribution of 5-clusters on frozen surfaces.}
\label{dist_5clust}
\end{figure}

\begin{figure}[ht]
\begin{center}
\epsfig{file=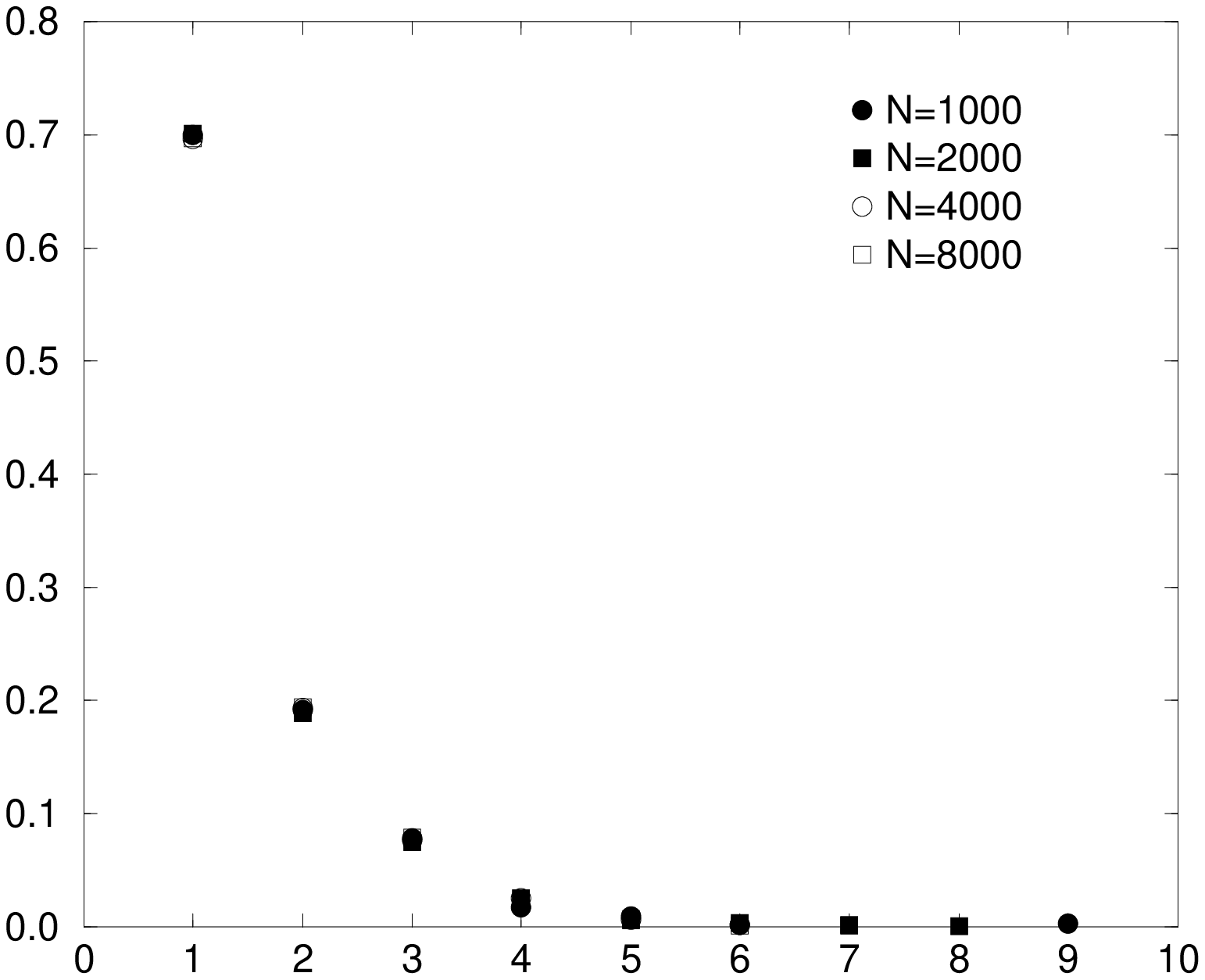,width=6.5cm,angle=0}
\end{center}
\caption{Size distribution of 7-clusters on frozen surfaces.}
\label{dist_7clust}
\end{figure}

In order to elucidate the role of flippers, it is necessary to have spatial informations on
surfaces when
removal of dipoles and energy freezing - at least apparently - occur.
So, it is very instructive to draw the surfaces. As can be seen in figure \ref{snapshots}, not only do
the defects of different kind - $i.e$ 5- and 7-vertices - separate, but, conversely, defects of the same
kind group together into clusters.
More precisely,
frozen surfaces look like a sea of 6-vertices with isolated clusters of defects
containing either 5-vertices or 7-vertices.
Figures \ref{dist_5clust} and \ref{dist_7clust} show the size distributions
of these clusters, measured on 16 independent frozen surfaces obtained after 1000 sweeps of thermalization
followed by sufficient
sweeps (up to 4.$10^{5}$) at zero temperature.
 It should be noticed that the sizes do not exceed few vertices. Moreover, the distribtution itself 
does not depend on
the size of the surfaces, meaning that the mechanism of freezing is local and
not influenced by  long range correlations.

\subsection{Flippers but no dipole creation}

The issue is now to understand the role of flippers on a surface
made of 5- and 7-clusters inside a sea of 6-vertices. There are two possibilities: \\
\noindent - 5-vertices and 7-vertices diffuse across the whole surface and, eventually,
some of them are close enough to form dipoles and allow decreasing of energy.
In this case, evolution could be very slow, but the system is not really frozen;

\noindent - 5-vertices and 7-vertices are moving but clusters are confined within spatial boundaries
so that no dipoles are created. In this case, the system is for ever
trapped in a set of metastable states.

Arguments suggesting that flippers cannot lead to dipole creation are now given.
The starting point is a state with clusters of different kinds (5-clusters and 7-clusters),
inside a sea of 6-vertices (about $85 \%$ of the surface).

The 5-clusters have the following property: {\it  if
there exists a convex contour} $\cal C${\it surrounding a
5-cluster and containing no 7-vertex, then, at zero temperature, $\cal C$  will never be crossed by a 5-vertex
coming from the cluster, i.e there will never be 5-vertices outside} ${\cal C}$. Convex contour means
that the number of links attached to each vertex of ${\cal C}$  and going
outside of ${\cal C}$ is larger or equal to the number of links going inside $\cal C$
- the links of ${\cal C}$ are not taken into account. The proof
runs as follows :\\
Let  a region $\cal R$, containing only  5- and 6-vertices,
be  bounded by a loop $\cal C$, made of links connecting 5- and/or 6-vertices.
It is supposed  that there are only 6-vertices outside $\cal C$. Then, moves of
type $a)$ are performed in $\cal R$ : \\
$\bullet$  The boundary ${\cal C}$ cannot be broken: consider
a link $l \in {\cal C}$. The only possibility for $l$ to be moved is to belong to type
$(IIIb)$ (flipper). In this case, because of the convexity of the contour, one of the two
neighbouring 5-vertices is inevitably outside $\cal C$. It would contradict
the starting hypothesis: so, $l$ cannot be moved.\\
$\bullet$ Consider now a set of four vertices $\{ \alpha_{1},\beta_{1},
\alpha_{2},\beta_{2}\}$ involved in a move of type $a)$, where  $\alpha_{1}$,
$\alpha_{2}$ are 5-vertices and $\beta_{1}$, $\beta_{2}$ are 6-vertices. As shown above, the link
$\beta_{1}$-$\beta_{2}$  cannot belong to $\cal C$. After the $T_1$ move,
$\alpha_{1}$, $\alpha_{2}$ are 6-vertices and $\beta_{1}$, $\beta_{2}$ are 5-vertices.
There are two possibilities that could lead to a diffusion of 5-vertices outside $\cal C$:\\
$(i)$ $\beta_{1}$ and $\beta_{2}$ are outside $\cal C$ or \\$(ii)$
$\beta_{1}$ or $\beta_{2}$ are outside $\cal C$.\\
Of course, the starting hypothesis prevents $\alpha_{1}$ and $\alpha_{2}$ to be
outside $\cal C$.  So, for the case $(i)$,
 $\cal C$ must contain $\alpha_{1}$ and $\alpha_{2}$,
but neither $\beta_{1}$ nor $\beta_{2}$. It is easy to see that such a configuration implies
a non convex boundary $\cal C$. For the case $(ii)$, $\cal C$ must contain $\alpha_{1}$,
$\alpha_{2}$ and $\beta_{1}$ (resp. $\beta_{2}$) but not  $\beta_{2}$ (resp. $\beta_{1}$).
It also implies a non convex boundary at $\beta_{1}$ (resp. $\beta_{2}$).

To conclude, diffusion of 5-vertices outside $\cal C$ is not allowed.

Consider now the 7-clusters. Unfortunately, last arguments given for 5-clusters do not work.
In particular, a convex boundary can be deformed by a flipper.
However, it is possible to have a rather reliable result. Consider
a 7-cluster and perform - by computer simulations -  $T_1$ moves of type $b)$ only in this cluster.
Then, the pattern of the cluster evolves through flippers but, as can be seen by drawing surfaces,
it does not extend much and never reaches any 5-cluster.
These simulations take very few computer time because 7-clusters contain only fews
vertices.
So, it is possible to perform very extensive runs and to have great confidence level in
this result.

To summarize, it is very likely  that the system gets trapped and follows
a constant energy walk in a set of states; these states are made of 5- and
7-clusters and are subject only to flippers that do not destroy the cluster
structure - in particular, flippers cannot create dipoles.

\section{Inherent structures and energy landscape}
The quenching procedure to zero temperature used in this work maps equilibrium 
states to frozen sets of metastable states. It is in some sense a steepest-descent 
dynamics to inherent states as defined by Stillinger and Weber \cite{StillWeber}.
This approach of glassy behaviours is based on a decomposition of the configuration 
space into basins, called inherent structures. Each basin contains the states
mapped through steepest-descent to the same local minimum in the energy 
landscape. This method provides a deeper understanding of glassy behaviours
in terms of fast intra-basins and slow inter-basins dynamics, with the definition of
a configurational entropy counting the number of inherent structures with a given 
energy or, alternatively, a given free energy \cite{Ritort1}. However, this symbolic dynamics
between inherent structures is not necessarily relevant for of all glassy systems \cite{Ritort2} \cite{Biroli}.

\begin{figure}[ht]
\begin{center}
\epsfig{file=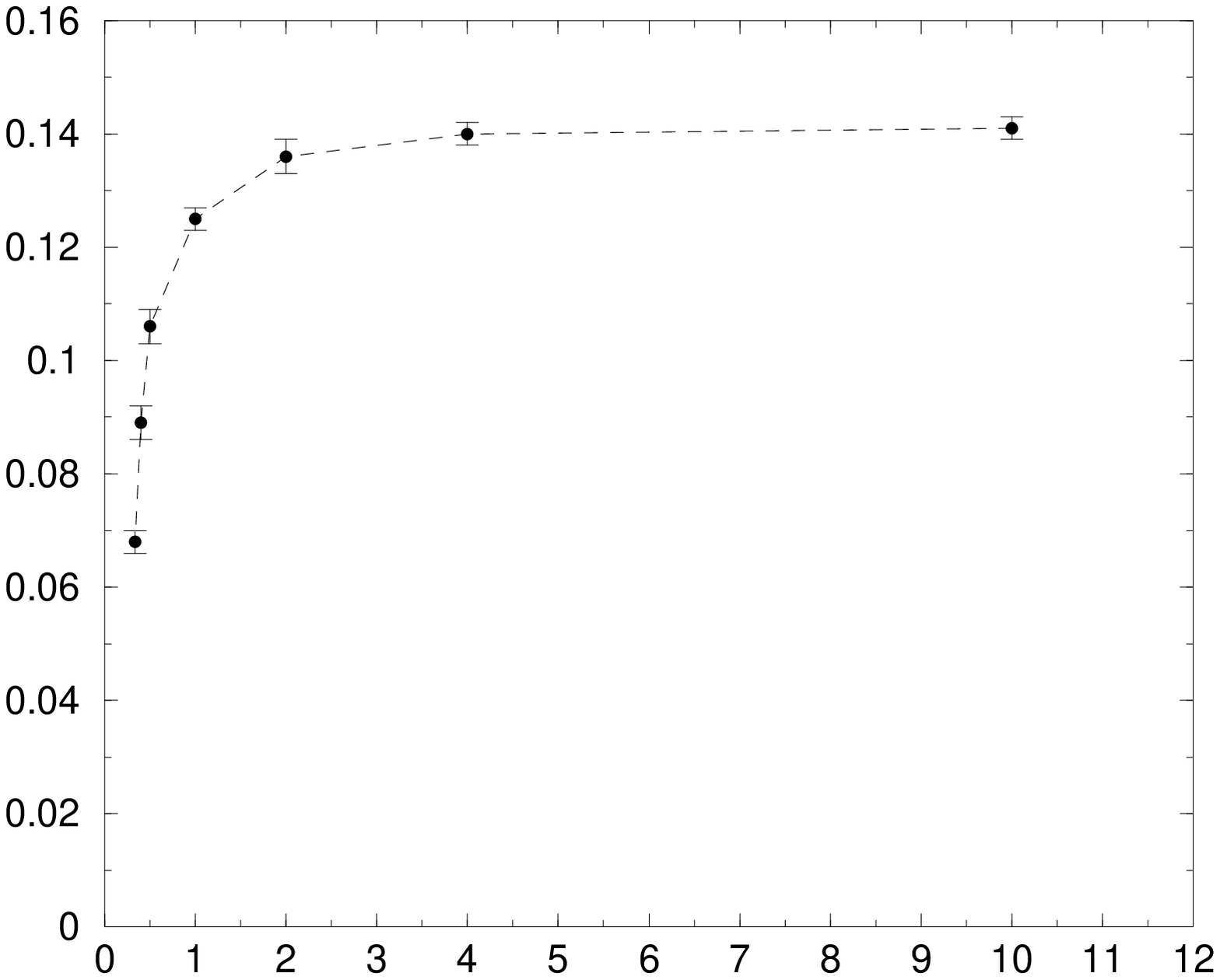,width=6.5cm,angle=0}
\end{center}
\caption{Frozen energy per site $E_{f}/N$ $versus$ temperature, for $N=400$.}
\label{finite}
\end{figure}

For the model studied in this paper, it would be very interesting to use 
the Stillinger and Weber approach. But, as the model is discrete, the zero
temperature dynamics is not a deterministic mapping to metastable states.
Moreover, these inherent structures are degenerate, as shown in the previous section.
So, it would require slight modifications to define unambiguously concepts like 
inherent structures, configurational entropy ... The present work is a first step in this approach 
as it provides a characterization of the metastable states reached after a quench from infinite 
temperature equilibrium states. To go further, the frozen
energy per site following a quench from finite temperature to zero temperature
has been calculated for rather small surfaces ($N=400$), as preliminary result. 
Figure \ref{finite} shows the variation of $E_{f}(T)/N$ with temperature.
 As can be seen, this quantity
is more or less independent of $T$ for $T \ge T_{o}$ with $T_{o} \simeq 2$, and strongly
decreases for $T < T_{o}$, meaning that the system explores deeper valleys in
the energy landscape.

\section{Conclusion}
The model of random surfaces with elastic curvature energy studied in this paper
shows rather trivial static properties. In particular, there is no phase
transition driven by curvature coupling constant - curvature energy simply
smoothes surfaces and the whole fractal structure of baby universes disappears
at low temperature equilibrium. 

However, the system reveals complex dynamical behaviour, though there is no
frustration nor quenched disorder. In particular, when
the system is quenched from totally disordered infinite temperature state
to low temperature, glassy behaviour takes place and relaxation time becomes
exponentially long as temperature is lowered.

Glassy behaviour is dramatically strengthened for deep quench at
zero temperature. In this case, simulations and some arguments strongly suggest
that the system really freezes: after a rapid decay,
the energy stops and the ground state is never reached. Equivalently,
freezing corresponds to defects (5- and 7-vertices) that cannot be removed
by the system.
This property does not depend on the size of the system
and should be still valid at the thermodynamical limit. In fact, even if energy
stops, there is a remaining dynamics of the system, consisting in flippers,{\it i.e}
$T_1$ moves that locally rotates defects of same kind. But flippers cannot make
energy evolve because of the spatial repartion of defects: they group together
into clusters made of one kind of defects . Then, flippers cannot lead to diffusion of defects
outside some bounded regions, and dipoles, that could lower energy, are not created.
The system is forever trapped
in a set of metastable states. Similar results are obtained for reaction-diffusion models \cite{drou_comm}. 
However, there is still an open question: why do defects
of same kind group together into clusters? In particular, it would be very interesting
to study the evolution of the fractal structure of baby universes during the deep
quench. At infinite temperature,
it is known that $d_{H}=4$. In frozen states, there are no more baby universes. But preliminary
results suggest that the fractal
dimension of frozen surfaces is significantly greater than 2 - the value for flat
surfaces. So, a guess can be made: frozen surfaces keep memory of their initial baby universes
structure through structure of clusters. The mechanism of clusters creation would be
the following : each initial baby universe is more or less isolated from the rest of the surface;
then, just after a quench to zero temperature, each baby universe tries to reach its own local 
ground state. But
in each baby universe, the difference between the number of 5-vertices and the number
of 7-vertices is topologically constrained by the boundary of the baby universe. So, after a rapid
multi-relaxation period where boundaries of baby universes are not expected to be much modified, 
some regions would be left
with an excess of 5-vertices and some others with an excess of 7-vertices. This mechanism could 
lead to creation of clusters. To summarize, clusters structure would be a trace
of initial baby universes structure.\\

It would be very interesting to use the inherent structures approach
of Stillinger and Weber for the system studied here, and to see wether this method can
provide a better understanding of the glassy behaviour observed at finite temperature. The present 
work is in some sense a first step in this direction.  

\section{Acknowledgements}
I am grateful to C. Godr\`eche for his help, suggestions and encouragments
all along this work. I would like to thank J-M Luck and J-M Drouffe for
interesting discussions and suggestions. I wish to acknowledge F. David for
helpfull discussions about strong coupling expansion of planar graphs.
Finally, I would like to thank the laboratory SPEC at CEA Saclay where
part of this work was done using computer facilities.

\end{document}